%% file: main.tex
\pdfoutput=1
\documentclass[type=preprint,linenumbering=off]{sphenix}

\usepackage{upgreek}
\usepackage[inline]{enumitem}
\usepackage{cite}
\usepackage{setspace}
\usepackage{cellspace}
\usepackage{amssymb}
\usepackage{adjustbox}
\usepackage{float}

\usepackage{orcidlink}

\begin{document}
\include{defs}

\title{Measurement of charged hadron multiplicity in \auau collisions at $\sqrt{\text{s}_{\text{NN}}}$\;=\;200~GeV with the \sphenix detector}
\author{\sphenix Collaboration\footnote{See the appendix for the list of collaboration members}}
\date{\today}

\doctag{sPH-BULK-2025-01}
\docdoi{10.1007/JHEP08(2025)075}

\maketitle

\begin{abstract}
  The pseudorapidity distribution of charged hadrons produced in \auau collisions at a center-of-mass energy of $\sqrtsnn=200\gev$ is measured using data collected by the \sphenix detector. Charged hadron yields are extracted by counting cluster pairs in the inner and outer layers of the Intermediate Silicon Tracker, with corrections applied for detector acceptance, reconstruction efficiency, combinatorial pairs, and contributions from secondary decays. The measured distributions cover $|\eta|<1.1$ across various centralities, and the average pseudorapidity density of charged hadrons at mid-rapidity is compared to predictions from Monte Carlo heavy-ion event generators. This result, featuring full azimuthal coverage at mid-rapidity, is consistent with previous experimental measurements at the Relativistic Heavy Ion Collider, thereby supporting the broader \sphenix physics program.
\end{abstract}

\input{introduction.tex}
\input{detector.tex}
\input{selection.tex}
\input{analysis.tex}
\input{systematic.tex}
\input{result.tex}
\input{conclusion.tex}

\section*{Acknowledgements}

We thank the Collider-Accelerator Division, SDCC and Physics
Departments at Brookhaven National Laboratory and the staff of the
other sPHENIX participating institutions for their vital
contributions. We acknowledge support from the Office of Nuclear
Physics and Graduate Student Research (SCGSR) program in the Office of
Science of the U.S. Department of Energy, the U.S. National Science
Foundation, the National Science and Technology Council, the Ministry
of Education of Taiwan, and the Ministry of Economic Affairs (Taiwan),
the Ministry of Education, Culture, Sports, Science, and Technology
and the Japan Society for the Promotion of Science (Japan), Basic
Science Research Programs through NRF funded by the Ministry of
Education and the Ministry of Science and ICT (Korea) and the Swedish
Research Council, VR (Sweden).

\clearpage

\bibliographystyle{unsrturl}
\bibliography{bib/references}

\newpage

\appendix

\section*{The sPHENIX Collaboration}

\begin{flushleft} 
\small

M.~I.~Abdulhamid$^{13}$,
U.~Acharya\,\orcidlink{0000-0001-8560-963X}\,$^{13}$,
E. R.~Adams$^{7}$,
G.~Adawi$^{13}$,
C.~A.~Aidala\,\orcidlink{0000-0001-9540-4988}\,$^{26}$,
Y.~Akiba$^{39}$,
M.~Alfred$^{14}$,
S.~Ali$^{13}$,
A.~Alsayegh$^{10}$,
S.~Altaf$^{15}$
% ...

\section*{Collaboration Institutes}

$^{1}$ Augustana University, Sioux Falls, South Dakota\\
$^{2}$ Baruch College, City University of New York, New York, New York\\
$^{3}$ Brookhaven National Laboratory, Upton, New York\\
$^{4}$ University of California, Los Angeles, California\\
$^{5}$ Central China Normal University, Wuhan, Hubei\\
$^{6}$ IRFU, CEA, Universit\'{e} Paris-Saclay, Gif-sur-Yvette, France\\
$^{7}$ University of Colorado, Boulder, Colorado\\
$^{8}$ Columbia University, New York, New York\\
$^{9}$ Debrecen University, Debrecen, Hungary\\
$^{10}$ Florida Agricultural and Mechanical University, Tallahassee, Florida\\
$^{11}$ Florida State University, Tallahassee, Florida\\
$^{12}$ Fudan University, Shanghai\\
$^{13}$ Georgia State University, Atlanta, Georgia\\
$^{14}$ Howard University, Washington, District of Columbia\\
$^{15}$ University of Illinois at Urbana-Champaign, Urbana, Illinois\\
$^{16}$ Institute for Information Industry, Taipei\\
$^{17}$ Iowa State University, Ames, Iowa\\
$^{18}$ Japan Atomic Energy Agency, Naka, Ibaraki, Japan\\
$^{19}$ Korea University, Seoul, Korea\\
$^{20}$ Lawrence Berkeley National Laboratory, Berkeley, California\\
$^{21}$ Lawrence Livermore National Laboratory, Livermore, California\\
$^{22}$ Lehigh University, Bethlehem, Pennsylvania\\
$^{23}$ Los Alamos National Laboratory, Los Alamos, New Mexico\\
$^{24}$ Lund University, Lund, Sweden\\
$^{25}$ Massachusetts Institute of Technology, Cambridge, Massachusetts\\
$^{26}$ University of Michigan, Ann Arbor, Michigan\\
$^{27}$ Morgan State University, Baltimore, Maryland\\
$^{28}$ Muhlenberg College, Allentown, Pennsylvania\\
$^{29}$ Nara Women's University, Nara, Nara, Japan\\
$^{30}$ National Central University, Taoyuan City\\
$^{31}$ National Research Nuclear University, MEPhI, Moscow Engineering Physics Institute, Moscow, Russia\\
$^{32}$ National Taiwan University, Taipei\\
$^{33}$ University of North Carolina, Greensboro, North Carolina\\
$^{34}$ Oak Ridge National Laboratory, Oak Ridge, Tennessee\\
$^{35}$ Ohio University, Athens, Ohio\\
$^{36}$ Peking University, Beijing\\
$^{37}$ Purdue University, West Lafayette, Indiana\\
$^{38}$ Pusan National University, Pusan, Korea\\
$^{39}$ RIKEN Nishina Center for Accelerator-Based Science, Wako, Saitama, Japan\\
$^{40}$ RIKEN BNL Research Center, Brookhaven National Laboratory, Upton, New York\\
$^{41}$ Rikkyo University, Toshima, Tokyo, Japan\\
$^{42}$ Rutgers University, Piscataway, New Jersey\\
$^{43}$ State University of New York, Stony Brook, New York\\
$^{44}$ Sun Yat-sen University, Guangzhou, Guangdong\\
$^{45}$ University of Tennessee, Knoxville, Tennessee\\
$^{46}$ University of Texas, Austin, Texas\\
$^{47}$ Vanderbilt University, Nashville, Tennessee\\
$^{48}$ Wayne State University, Detroit, Michigan\\
$^{49}$ Weizmann Institute of Science, Rehovot, Israel\\
$^{50}$ Yonsei University, Seoul, Korea\\
$^{51}$ United States Merchant Marine Academy, Kings Point, New York\\

\end{flushleft}

\end{document}

%% file: defs.tex
\newcommand{\dndeta}{\mbox{d$N_{\rm ch}$/d$\eta$}\xspace}

\newcommand{\sphenix}{\mbox{sPHENIX}\xspace}
\newcommand{\phenix}{\mbox{PHENIX}\xspace}
\newcommand{\phobos}{\mbox{PHOBOS}\xspace}
\newcommand{\brahms}{\mbox{BRAHMS}\xspace}
\newcommand{\alice}{\mbox{ALICE}\xspace}
\newcommand{\cms}{\mbox{CMS}\xspace}
\newcommand{\atlas}{\mbox{ATLAS}\xspace}
\newcommand{\starexp}{\mbox{STAR}\xspace}

\newcommand{\mvtx}{\mbox{MVTX}\xspace}
\newcommand{\intt}{\mbox{INTT}\xspace}
\newcommand{\tpc}{\mbox{TPC}\xspace}
\newcommand{\tpot}{\mbox{TPOT}\xspace}
\newcommand{\emcal}{\mbox{EMCAL}\xspace}
\newcommand{\ihcal}{\mbox{iHCAL}\xspace}
\newcommand{\ohcal}{\mbox{oHCAL}\xspace}
\newcommand{\mbd}{\mbox{MBD}\xspace}
\newcommand{\zdc}{\mbox{ZDC}\xspace}
\newcommand{\sepd}{\mbox{sEPD}\xspace}
\newcommand{\smd}{\mbox{SMD}\xspace}

\newcommand{\minbias}{\textsc{Min. Bias}\xspace}

\newcommand{\hic}{\mbox{$A$$+$$A$}\xspace}
\newcommand{\ee}{\mbox{$e^{+}$$+$$e^{-}$}\xspace}
\newcommand{\ep}{\mbox{$e$$+$$p$}\xspace}
\newcommand{\eA}{\mbox{$e$$+$$A$}\xspace}
\newcommand{\pA}{\mbox{$p$$+$$A$}\xspace}
\newcommand{\AuAu}{\mbox{Au$+$Au}\xspace}
\newcommand{\uu}{\mbox{U$+$U}\xspace}
\newcommand{\apa}{\mbox{A$+$A}\xspace}
\newcommand{\auau}{\mbox{Au$+$Au}\xspace} 
\newcommand{\alal}{\mbox{Al$+$Al}\xspace} 
\newcommand{\oo}{\mbox{O$+$O}\xspace} 
\newcommand{\agag}{\mbox{Ag$+$Ag}\xspace} 
\newcommand{\cucu}{\mbox{Cu$+$Cu}\xspace} 
\newcommand{\xexe}{\mbox{Xe$+$Xe}\xspace} 
\newcommand{\raa}{\mbox{$R_{AA}$}\xspace}
\newcommand{\pbpb}{\mbox{Pb$+$Pb}\xspace} 
\newcommand{\ppb}{\mbox{p$+$Pb}\xspace} 
\newcommand{\pdau}{\mbox{$p(d)$$+$Au}\xspace} 
\newcommand{\aj}{\mbox{$A_J$}\xspace} 
\newcommand {\pp}{\mbox{$p$$+$$p$}\xspace}
\newcommand {\ppbar}{\mbox{$p$$+$$\overline{p}$}\xspace}
\newcommand{\pT}{\mbox{${p_T}$}\xspace}
\newcommand{\jpsi}{\mbox{$J/\psi$}\xspace}
\newcommand{\sqrts}{\mbox{$\sqrt{s}$}\xspace}
\newcommand{\sqrtsnn}{\mbox{$\sqrt{s_{\scriptscriptstyle NN}}$}\xspace}
\newcommand{\npart}{$N_{\mathrm{part}}$\xspace}
\newcommand{\ncoll}{$N_{\mathrm{coll}}$\xspace}
\newcommand{\qgp}{\mbox{quark-gluon plasma}\xspace}
\newcommand{\jt}{\mbox{$J_T$}} \newcommand{\qhat}{\mbox{$\hat{q}$}\xspace}
\newcommand{\Qsqr}{\mbox{$Q^2$}} \newcommand{\CuCu}{\mbox{Cu$+$Cu}\xspace}
\newcommand{\PbPb}{\mbox{Pb$+$Pb}\xspace} 
\newcommand{\pPb}{\mbox{$p$$+$Pb}\xspace}
\newcommand{\gjet}{\mbox{$\gamma$-jet}\xspace}
\newcommand{\Qmax}{\mbox{$Q_{\max}$}} \newcommand{\ET}{\mbox{$E_T$}}
\newcommand{\Et}{\mbox{$E_T$}} \newcommand{\kt}{\mbox{$k_T$}}
\newcommand{\RAA}{\mbox{$R_{AA}$}\xspace} \newcommand{\IAA}{\mbox{$I_{AA}$}}
\newcommand{\pt}{\mbox{${p_{\mathrm{T}}}$}\xspace}
\newcommand{\nb}{\mbox{nb$^{-1}$}\xspace}
\newcommand{\pb}{\mbox{pb$^{-1}$}\xspace}
\newcommand{\fb}{\mbox{fb$^{-1}$}\xspace}

\newcommand{\Dzero}{$D^{0}$\xspace}
\newcommand{\dTokpi}{$D^{0}\rightarrow K^-\pi^+$\xspace}
\newcommand{\mkpi}{$m_{K^-\pi^+}$\xspace}
\newcommand {\bbbar}{\mbox{$b\overline{b}$}\xspace}
\newcommand {\ccbar}{\mbox{$c\overline{c}$}\xspace}

\newcommand{\highpt}{high-${\rm p_{_{T}}}$}
\newcommand{\lessim}{{\stackrel{<}{\sim}}} \newcommand{\eqnpt}{p_T}
\newcommand{\dAu}{\mbox{$d$$+$Au}\xspace}
\newcommand{\pAu}{\mbox{$p$$+$Au}\xspace}
\newcommand{\pau}{\mbox{$p$$+$Au}\xspace}
\newcommand{\pAl}{\mbox{$p$$+$Al}\xspace}
\newcommand{\gevsq}{\mbox{${\rm~GeV}^2$}\xspace}
\newcommand{\fastjet}{\mbox{\sc FastJet}\xspace}
\newcommand{\hepmc}{\mbox{\sc HepMC2}\xspace}
\newcommand{\geant}{\mbox{\sc Geant4}\xspace}
\newcommand{\antikt}{\mbox{anti-$k_T$}\xspace}
\newcommand{\pythia}{\mbox{\sc Pythia8}\xspace}
\newcommand{\funforall}{\mbox{\sc Fun4All}\xspace}
\newcommand{\kfparticle}{\mbox{\sc KFParticle}\xspace}
\newcommand{\decayfinder}{\mbox{\sc DecayFinder}\xspace}
\newcommand{\hftrackeff}{\mbox{\sc HFTrackEfficiency}\xspace}
\newcommand{\rapgap}{\mbox{\sc Rapgap}\xspace}
\newcommand{\milou}{\mbox{\sc Milou}\xspace}
\newcommand{\pyquen}{\mbox{\tt Pyquen}\xspace}
\newcommand{\hijing}{\mbox{\tt HIJING}\xspace}
\newcommand{\ampt}{\mbox{\tt AMPT}\xspace}
\newcommand{\epos}{\mbox{\tt EPOS4}\xspace}
\newcommand{\jewel}{\mbox{\tt Jewel}\xspace}
\newcommand{\roofit}{\mbox{\sc RooFit}\xspace}
\newcommand{\roounfold}{\mbox{\sc RooUnfold}\xspace}
\newcommand{\beetle}{\mbox{\sc Beetle}\xspace}
\newcommand{\gj}{\mbox{$\gamma$+jet}\xspace}
\newcommand{\gh}{\mbox{$\gamma$+hadron}\xspace}
\newcommand{\martinimusic}{\mbox{\sc Martini+Music}\xspace}
\newcommand{\martini}{\mbox{\sc Martini}\xspace}
\newcommand{\music}{\mbox{\sc Music}\xspace}
\newcommand{\Ephenix}{Electron-Ion Collider (EIC) detector built
  around the BaBar magnet and sPHENIX calorimetry\xspace}
\newcommand{\ephenix}{EIC detector built around the BaBar magnet and
  sPHENIX calorimetry\xspace} 
\newcommand{\refdesign}{reference design\xspace}
\newcommand{\refconfig}{reference configuration\xspace}
\newcommand{\dijet}{\mbox{dijet}\xspace}
\newcommand{\fake}{\mbox{fake}\xspace}
\newcommand{\fast}{\mbox{fast}\xspace}
\newcommand{\veryfast}{\mbox{very fast}\xspace}
\newcommand{\epem}{\mbox{$e^+e^-$}\xspace}
\newcommand{\onewidth}{0.6\linewidth}
\newcommand{\twowidth}{0.48\linewidth}
\newcommand{\threewidth}{0.32\linewidth}

\newcommand{\egoing}{\mbox{electron-going}\xspace}
\newcommand{\hgoing}{\mbox{hadron-going}\xspace}
\newcommand{\egodir}{electron-going direction\xspace}
\newcommand{\hgodir}{hadron-going direction\xspace}
\newcommand{\bigcell}[2]{\begin{tabular}{@{}#1@{}}#2\end{tabular}}

\def\sPlot{\mbox{\em sPlot}\xspace}
\def\sPlots{\mbox{\em sPlots}\xspace}
\def\sWeights{\mbox{\em sWeights}\xspace}
\newcommand{\unit}[1]{\ensuremath{\rm\,#1}\xspace}          

\newcommand{\tev}{\ensuremath{\mathrm{\,Te\kern -0.1em V}}\xspace}
\newcommand{\gev}{\ensuremath{\mathrm{\,Ge\kern -0.1em V}}\xspace}
\newcommand{\mev}{\ensuremath{\mathrm{\,Me\kern -0.1em V}}\xspace}
\newcommand{\kev}{\ensuremath{\mathrm{\,ke\kern -0.1em V}}\xspace}
\newcommand{\ev}{\ensuremath{\mathrm{\,e\kern -0.1em V}}\xspace}
\newcommand{\gevc}{\ensuremath{{\mathrm{\,Ge\kern -0.1em V\!/}c}}\xspace}
\newcommand{\mevc}{\ensuremath{{\mathrm{\,Me\kern -0.1em V\!/}c}}\xspace}
\newcommand{\gevcc}{\ensuremath{{\mathrm{\,Ge\kern -0.1em V\!/}c^2}}\xspace}
\newcommand{\gevgevcccc}{\ensuremath{{\mathrm{\,Ge\kern -0.1em V^2\!/}c^4}}\xspace}
\newcommand{\mevcc}{\ensuremath{{\mathrm{\,Me\kern -0.1em V\!/}c^2}}\xspace}

\def\km   {\ensuremath{\rm \,km}\xspace}
\def\m    {\ensuremath{\rm \,m}\xspace}
\def\cm   {\ensuremath{\rm \,cm}\xspace}
\def\cma  {\ensuremath{{\rm \,cm}^2}\xspace}
\def\mm   {\ensuremath{\rm \,mm}\xspace}
\def\mma  {\ensuremath{{\rm \,mm}^2}\xspace}
\def\mum{\ensuremath{\upmu\mathrm{m}}\xspace}
\def\muma {\ensuremath{\rm \,\upmu\mathrm{m}^2}\xspace}
\def\nm   {\ensuremath{\rm \,nm}\xspace}
\def\fm   {\ensuremath{\rm \,fm}\xspace}
\def\barn{\ensuremath{\rm \,b}\xspace}
\def\barnhyph{\ensuremath{\rm -b}\xspace}
\def\mbarn{\ensuremath{\rm \,mb}\xspace}
\def\mub{\rm \,\textmu b\xspace}
\def\mbarnhyph{\ensuremath{\rm -mb}\xspace}
\def\nb {\ensuremath{\rm \,nb}\xspace}
\def\invnb {\ensuremath{\mbox{\,nb}^{-1}}\xspace}
\def\pb {\ensuremath{\rm \,pb}\xspace}
\def\invpb {\ensuremath{\mbox{\,pb}^{-1}}\xspace}
\def\fb   {\ensuremath{\mbox{\,fb}}\xspace}
\def\invfb   {\ensuremath{\mbox{\,fb}^{-1}}\xspace}
\def\khz   {\ensuremath{\mbox{\,kHz}}\xspace}
\def\mhz   {\ensuremath{\mbox{\,MHz}}\xspace}
\def\ghz   {\ensuremath{\mbox{\,GHz}}\xspace}
\def\mbs   {\ensuremath{\mbox{\,Mb/s}}\xspace}
\def\gbs   {\ensuremath{\mbox{\,Gb/s}}\xspace}
\def\tbs   {\ensuremath{\mbox{\,Tb/s}}\xspace}

\def\sec  {\ensuremath{\rm {\,s}}\xspace}
\def\ms   {\ensuremath{{\rm \,ms}}\xspace}
\def\mus  {\rm \,\textmu s\xspace}
\def\ns   {\ensuremath{{\rm \,ns}}\xspace}
\def\ps   {\ensuremath{{\rm \,ps}}\xspace}
\def\fs   {\ensuremath{\rm \,fs}\xspace}

\def\mhz  {\ensuremath{{\rm \,MHz}}\xspace}
\def\khz  {\ensuremath{{\rm \,kHz}}\xspace}
\def\hz   {\ensuremath{{\rm \,Hz}}\xspace}

\def\invps{\ensuremath{{\rm \,ps^{-1}}}\xspace}

\def\yr   {\ensuremath{\rm \,yr}\xspace}
\def\hr   {\ensuremath{\rm \,hr}\xspace}
\def\degc {\ensuremath{^\circ}{C}\xspace}
\def\degk {\ensuremath {\rm K}\xspace}

\def\Xrad {\ensuremath{X_0}\xspace}
\def\NIL{\ensuremath{\lambda_{int}}\xspace}
\def\mip {MIP\xspace}
\def\neutroneq {\ensuremath{\rm \,n_{eq}}\xspace}
\def\neqcmcm {\ensuremath{\rm \,n_{eq} / cm^2}\xspace}
\def\kRad {\ensuremath{\rm \,kRad}\xspace}
\def\MRad {\ensuremath{\rm \,MRad}\xspace}
\def\ci {\ensuremath{\rm \,Ci}\xspace}
\def\mci {\ensuremath{\rm \,mCi}\xspace}

\def\sx    {\ensuremath{\sigma_x}\xspace}    
\def\sy    {\ensuremath{\sigma_y}\xspace}   
\def\sz    {\ensuremath{\sigma_z}\xspace}    

\newcommand{\stat}{\ensuremath{\mathrm{(stat)}}\xspace}
\newcommand{\syst}{\ensuremath{\mathrm{(syst)}}\xspace}
\newcommand{\model}{\ensuremath{\mathrm{(model)}}\xspace}

\newcommand{\ten}[1]{\ensuremath{\times 10^{#1}}}
\def\order{{\ensuremath{\cal O}}\xspace}
\newcommand{\chisq}{\ensuremath{\chi^2}\xspace}
\newcommand{\erfc}[1]{\ensuremath{\rm{Erfc}(#1)}\xspace}

\def\deriv {\ensuremath{\mathrm{d}}}

\def\gsim{{~\raise.15em\hbox{$>$}\kern-.85em
          \lower.35em\hbox{$\sim$}~}\xspace}
\def\lsim{{~\raise.15em\hbox{$<$}\kern-.85em
          \lower.35em\hbox{$\sim$}~}\xspace}

\newcommand{\DR}{\ensuremath{\Delta \text{R}}}
\newcommand{\DPhi}{\ensuremath{\Delta\phi}}
\newcommand{\DEta}{\ensuremath{\Delta\eta}}

%% file: introduction.tex
\section{Introduction}
\label{sec:introduction}
A hot medium of strongly interacting, deconfined quarks and gluons, known as the quark-gluon plasma (QGP), is formed in ultra-relativistic heavy-ion collisions~\cite{Busza:2018rrf}. The multiplicity and pseudorapidity ($\eta$) distributions of charged particles produced in these collisions are critical observables for characterizing the initial conditions and the subsequent hydrodynamic evolution of the QGP~\cite{Romatschke:2017ejr}. The dependence of charged-particle multiplicity on the colliding system, center-of-mass energy, and collision geometry provides several insights, such as into nuclear shadowing and gluon saturation effects~\cite{Albacete:2014fwa}. It can further reveal the relative contributions to particle production from hard scattering and soft processes~\cite{Kharzeev:2000ph, dEnterria:2011twh}. Studying the charged hadron multiplicity and its dependence on $\eta$ is essential for understanding the formation and properties of the QGP in heavy-ion collisions. 

At the Relativistic Heavy Ion Collider (RHIC)~\cite{HARRISON2003235}, measurements of the system-size dependence of charged-particle $\eta$ density, denoted as $\dndeta$, have been performed for diverse collision systems, such as copper-copper, gold-gold (\auau), and deuteron-gold collisions at various center-of-mass energies~\cite{PHENIX:2000owy,PHENIX:2004vdg,PHENIX:2018hho,PHOBOS:2000wxz,PHOBOS:2003fjw,PHOBOS:2010eyu,BRAHMS:2001gci,BRAHMS:2001llo,STAR:2001eyo}. Similarly, the \alice, \atlas, and \cms experiments at the Large Hadron Collider (LHC) have reported $\dndeta$ at mid-rapidity for proton-proton, lead-lead (\pbpb), proton-lead, and xenon-xenon (\xexe) collisions at \tev energy scales~\cite{ALICE:2009wpl,ALICE:2010cin,ALICE:2010khr,ALICE:2010mty,ALICE:2015juo,ALICE:2015olq,ALICE:2015qqj,ALICE:2017pcy,ALICE:2018cpu,ALICE:2020swj,CMS:2010wcx,CMS:2010qvf,CMS:2010tjh, CMS:2011aqh,CMS:2014kix,CMS:2015zrm,CMS:2017shj,CMS:2019gzk,CMS:2024ykx,ATLAS:2011ag,ATLAS:2015hkr}. These measurements have revealed several key empirical trends:
\begin{enumerate*}[label=(\arabic*)]
\item charged-particle production approximately follows a power-law scaling with center-of-mass energy;
\item central heavy ion collisions show a steeper increase in $\langle\dndeta\rangle$ as a function of center-of-mass energy compared to proton-proton and proton-nucleus collisions;
\item the values of \dndeta, normalised by the number of participating nucleons, \npart, have a non-linear increase with \npart;
\item the shapes of the \npart dependence remain consistent across different collision energies.
\end{enumerate*}
These findings provide an opportunity to test scaling laws and models tuned to data from different energy regimes and evaluate their applicability to other collision systems.

This analysis uses data collected by \sphenix~\cite{PHENIX:2015siv, Belmont:2023fau} during the RHIC \auau run at a nucleon-nucleon center-of-mass energy of $\sqrtsnn=200\gev$, with a beam-beam crossing angle of 2 milliradians, taken in October 2024 under zero magnetic field conditions. The charged hadron yield per unit $\eta$, corrected for detector acceptance, reconstruction efficiency, combinatorial pairs, and contributions from secondary decays, is determined by counting tracklets within $|\eta|<1.1$. Tracklets are reconstructed by pairing clusters from the inner and outer layers of the Intermediate Silicon Tracker (\intt) that point back to the primary interaction vertex. Results are reported as a function of $\eta$ for different \auau centrality classes. To further explore particle production mechanisms, \dndeta values at mid-rapidity, scaled by number of participant nucleon pairs \npart/2, are studied as a function of \npart. These measurements are compared to previous RHIC results and to predictions from the heavy-ion event generators \hijing~\cite{Wang:1991hta}, \ampt~\cite{Lin:2004en}, and \epos~\cite{Pierog:2013ria}.

%% file: detector.tex
\section{\sphenix Detector}
\label{sec:detector}
\sphenix is designed to be a general-purpose detector, aiming to measure jet and heavy-flavor probes of the QGP created in \auau collisions at RHIC. A four-component precision tracking system enables measurements of heavy flavor and jet substructure observables, while the electromagnetic and hadronic calorimeter system is crucial for measuring the energy of jets and identifying direct photons and electrons. 
This analysis is based on data collected using the \intt, the Minimum Bias Detectors (\mbd), and the Zero Degree Calorimeters (\zdc), which are described in detail.

The \intt is a two-layer barrel strip tracker with a clamshell geometry~\cite{INTT_ladder_NIM_cite}. Each layer comprises two sensor rings, offset in $\phi$ to provide full azimuthal coverage, and is situated between the MAPS-based Vertex Detector (MVTX) and the Time Projection Chamber in \sphenix.
It is designed to provide the \sphenix tracking system with the capability to associate reconstructed tracks to individual RHIC bunch crossings, enabling effective out-of-time pileup discrimination and suppression. 
The detector consists of 56 silicon ladders, 24 in the inner and 32 in the outer barrel, evenly spaced in a cylindrical configuration around the beam pipe at radial positions of approximately 7.2, 7.8, 9.7, and 10.3\cm from the beam axis. The active area of each ladder measures 45.6\cm in length and 2\cm in width and is composed of four silicon sensors. This design ensures hermetic coverage over $2\pi$ in azimuth and $|\eta|<1.1$ in pseudorapidity for collision vertices within $\pm10\cm$ of the nominal interaction point along the beam axis. The silicon sensors of each ladder are segmented into 52 silicon blocks, with each block containing 128 strips. The strips have a pitch of 78\mum, and extend 16 or 20\mm along the beam axis, resulting in a total of 6,656 readout channels per ladder. To minimize material interference and preserve track reconstruction accuracy, each silicon ladder is designed with a radiation length \ensuremath{X/X_0} of 1.14\%. Before the \intt, toward the collision point, the material budget is approximately 3.8\% \ensuremath{X/X_0} due to contributions from the beam pipe (1.8\% \ensuremath{X/X_0}) and the MVTX detector (2.0\% \ensuremath{X/X_0}). The single-hit detection efficiency of the \intt ladder is measured to be better than 99\% in test beam. Table~\ref{tab:INTT_spec} summarizes key specifications of the \intt.



\renewcommand{\arraystretch}{1.3}
\begin{table}[t!]
    \centering
    \caption{Specification of the \sphenix Intermediate Silicon Tracker (\intt).}
    \begin{tabular}{l c}
        \toprule[1pt]
        \textbf{Element} & \textbf{Value} \\
        \midrule[1pt]
        Number of ladders & 56 \\
        \multirow{2}{*}{Barrel radial distance to the beam line}  & 7.2 and 7.8\cm (Inner layer) \\
         & 9.7 and 10.3\cm (Outer layer)\\
        Radiation length (per ladder) & 1.14\% [\ensuremath{X/X_0}] \\
        Active area (per ladder) & $45.6 \times 2 \cma$  \\
        Number of channels (per ladder) & 6,656 \\
        Channel strip pitch & 78 \mum \\
        \multirow{2}{*}{Channel strip length} & 16\mm (strips within $\pm$13.0\cm along the beam axis) \\
         & 20\mm (otherwise) \\
        \bottomrule[1pt]
    \end{tabular}
    \label{tab:INTT_spec}
\end{table}
\renewcommand{\arraystretch}{1}

The \mbd is located on both sides of the interaction point at $\pm250\cm$ along the beam axis, covering the pseudorapidity range $3.51 < |\eta| < 4.61$~\cite{PHENIX:2003tlh}. Each side consists of 64 photomultiplier tubes (PMTs). The PMTs are arranged in three concentric rings around the beam pipe, ensuring full $2\pi$ azimuthal coverage. The \mbd serves as the primary detector for triggering minimum-bias (MB) events in heavy-ion collisions and provides essential global event information, including the event centrality and the measurement of the collision z-vertex position.

The \zdc of sPHENIX is located at a distance of approximately 18\m from both sides of the interaction point~\cite{Adler:2000bd}. It is a longitudinally segmented sampling hadronic calorimeter constructed from a tungsten alloy and optical fibers. These fibers transmit Cherenkov light produced by secondary charged particles within hadronic showers to PMTs. The \zdc is calibrated so that the single-neutron peak is positioned at the nominal value of 100 GeV. The ZDC is used in the determination of the MB criteria to distinguish \mbd-triggered events from beam-induced background.

%% file: selection.tex
\section{Data Selection}
\begin{figure}[t!]
  \centering
  \includegraphics[width=0.5\textwidth]{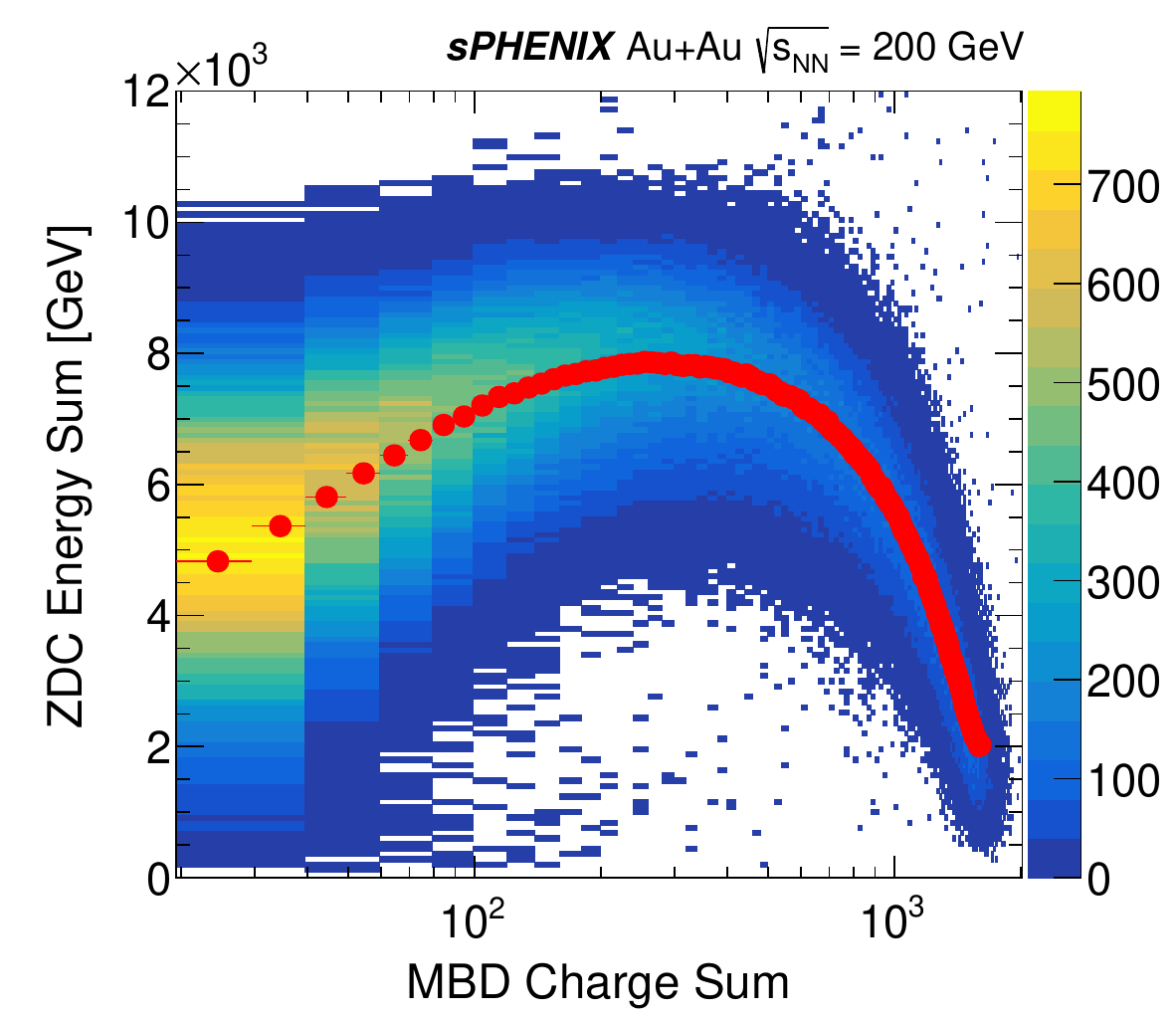}
  \caption{The correlation between the total energy in ZDCs and the \mbd charge sum is shown. The \mbd charge sum is in units of calibrated MIPs. The red points indicate the average \zdc energy as a function of \mbd charge sum.} 
  \label{fig:mbdzdc}
\end{figure}

Collision events were accepted using a hardware trigger which required at least two PMTs fired on each side of the \mbd. In the offline analysis, a set of MB selection criteria has been applied using signals from both the \mbd and \zdc to remove events consistent with beam-related backgrounds and non-hadronic interactions. Figure~\ref{fig:mbdzdc} shows the correlation of energy in the ZDCs to the \mbd charge sum for events passing the MB selection criteria. The \mbd charge sum is the sum of calibrated charge from all PMTs on both sides of the \mbd, and is calibrated such that one MIP is set at unity. A characteristic decrease in the total \zdc energy is observed at both high and low \mbd charge, corresponding to very central collisions (where only a small number of spectator neutrons deposit energy in the \zdc) and to peripheral collisions (where most spectator neutrons are bound in larger nuclear fragments and are thus deflected away from the \zdc by the RHIC magnets), respectively. This analysis utilizes 4.38\ten{6} MB events.

\begin{figure}[t!]
  \centering
  \includegraphics[width=0.5\textwidth]{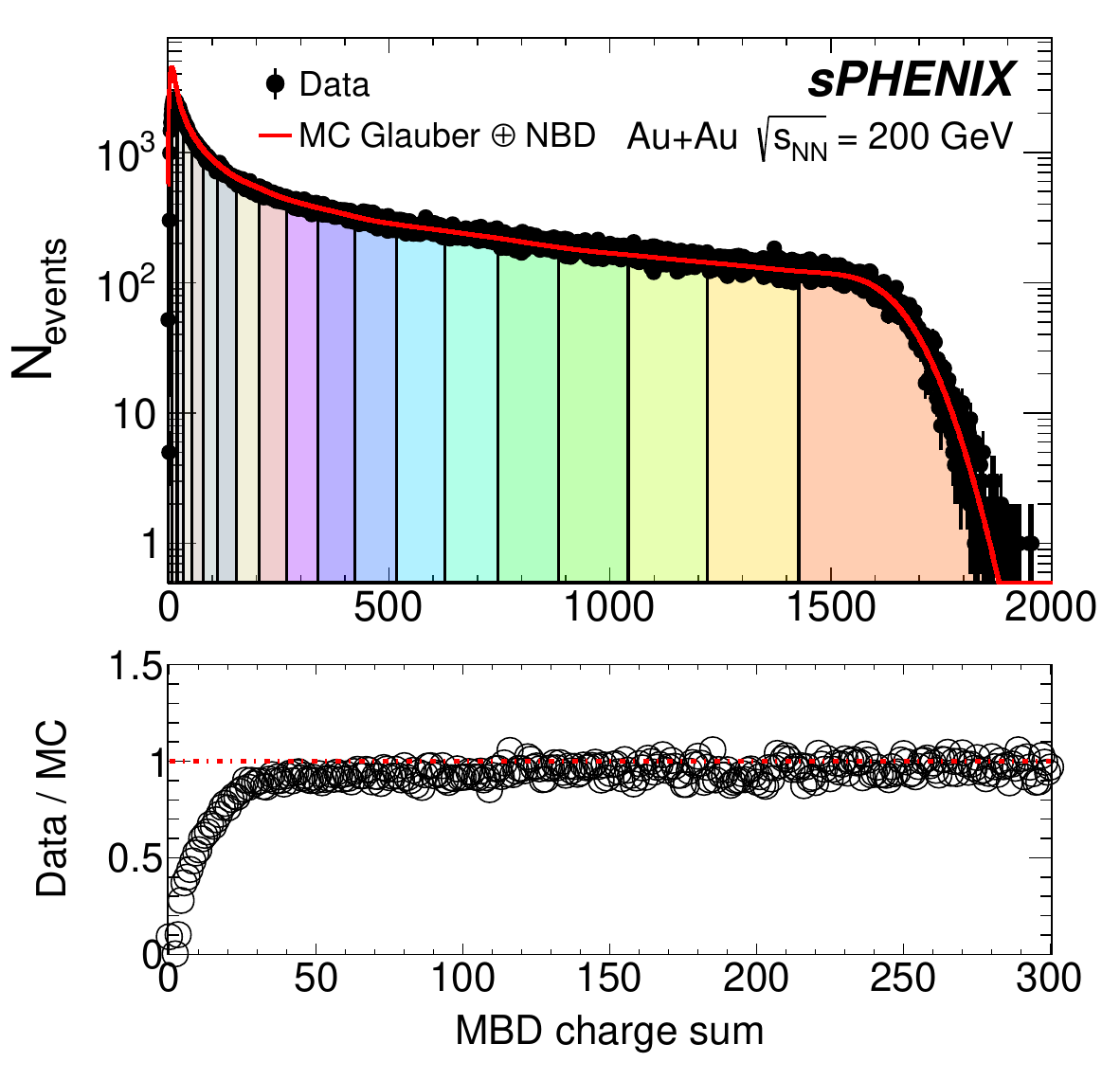}
  \caption{Top panel: the distribution of total charge deposited in the \mbd (solid black circles), the best fit of an MC Glauber model convolved with an NBD (red line), and centrality intervals for the MB events of \auau collisions at \sqrtsnn=200\gev collected with zero magnetic field (shaded regions). Bottom panel: the trigger efficiency as a function of the \mbd charge deposit in the peripheral collision region.} 
  \label{fig:centralityfiv}
\end{figure}

The \mbd charge sum is used in each event to assign a centrality value that characterizes the level of geometric overlap between the two colliding nuclei. Centrality percentiles are derived by fitting the \mbd charge sum distribution for all MB events to a model of the particle production and event sampling based on the convolution of a Monte Carlo (MC) Glauber simulation~\cite{Loizides:2014vua} with a negative binomial distribution (MC Glauber $\oplus$ NBD)~\cite{PHENIX:2004vdg,PHENIX:2013jxf}. Figure~\ref{fig:centralityfiv} shows the \mbd charge sum distribution, along with the best fit of the MC Glauber $\oplus$ NBD model and resulting centrality intervals. The bottom panel of Figure~\ref{fig:centralityfiv} shows the peripheral event region to highlight the MB trigger efficiency turn-on region. The total MB selection efficiency for inelastic Au+Au events is then determined by comparing the integral of the data distribution to the integral of the MC Glauber $\oplus$ NBD, normalized for \mbd charge sum $>150$. The resulting efficiency is $92.0\%^{+3.4\%}_{-3.1\%} (\text{syst.)}$, similar to previous \auau data-taking in \phenix where the \mbd was in a different location~\cite{PHENIX:2004vdg}. Glauber model parameters for \auau collisions at $\sqrtsnn=200\gev$ are shown in Table~\ref{tab:glauber}. 

\begin{table}[H]
    \centering
    \caption{Glauber model parameters for \auau collisions at $\sqrtsnn=200\gev$.}
    \begin{tabular}{l c}
        \toprule[1pt]
        \textbf{Glauber model parameter} & \textbf{Value} \\
        \midrule[1pt]
         $\sigma_{NN}$ [mb] & $42\pm3$ \\[2mm]
         Nuclear radius [fm] & $6.38^{+0.27}_{-0.13}$ \\[2mm]
         Skin depth [fm] & $0.535^{+0.020}_{-0.010}$ \\
         \bottomrule[1pt]
    \end{tabular}
    \label{tab:glauber}
\end{table}

A z-vertex cut, $|z_\mathrm{vtx}|\leqslant10\cm$ determined by the \intt, is imposed, ensuring the collisions take place in the position where the designed acceptance of the \intt can be maintained.
The \intt z-vertex is determined using tracklets, as described in Section~\ref{sec:analysis}. 

Three different MC event generators, \hijing, \ampt, and \epos, are used in this analysis to determine correction factors for extracting \dndeta from the \sphenix \intt measurement. \hijing, a Heavy Ion Jet Interaction Generator, is used to obtain the nominal results, while \ampt and \epos serve as alternative models to assess the uncertainty in the corrections due to differences in physics modeling.
\hijing models initial parton scatterings using perturbative QCD, describes jet fragmentation and hadronization with the Lund string model~\cite{ANDERSSON1987289}, and applies the Dual Parton Model~\cite{Capella:1992yb} to simulate soft interactions. It also incorporates effects such as jet quenching and nuclear shadowing in parton distribution functions. A Multi-Phase Transport model (\ampt) includes explicit interactions between initial minijet partons and final-state hadronic interactions. It extends \hijing by evolving initial partons with Zhang's parton cascade (ZPC) procedure~\cite{Zhang:1997ej} and the ART model~\cite{Li:1995pra} for the last stage of parton hadronization. \epos, based on Gribov-Regge theory~\cite{gribov1968reggeon, Drescher:2000ha}, includes nuclear effects such as Cronin transverse momentum broadening, parton saturation, screening, as well as high-density effects that lead to collective hadronization in hadron-hadron scattering.

The detector response is simulated using \geant~\cite{Geant4_cite} and processed through the same event reconstruction chain as the data. The \intt geometry in the detector simulation is adjusted based on survey measurements conducted after the \intt was installed in its current position within \sphenix. Simulations are generated using a vertex distribution with the mean and width set to match those observed in the collision data.

In line with previous measurements at RHIC and LHC~\cite{ALICE-PUBLIC-2017-005}, the primary charged hadrons are defined as prompt charged hadrons produced directly in the strong interaction and decay products of particles with proper decay length $c\tau < 1 \cm$, where $c$ is the speed of light in vacuum and $\tau$ is the proper lifetime of the particle. This definition excludes contributions from leptons, decay products of particles with longer lifetimes (such as $K_{S}^{0}$, $\Lambda$), and secondary interactions.

%% file: analysis.tex
\section{Analysis}
\label{sec:analysis}
The charged particle multiplicity is determined by counting tracklets, formed by pairing clusters with a small angular separation from two \intt layers. 
Two analysis approaches have been developed. The first closely follows the tracklet reconstruction methods outlined in the \phobos and \phenix  publications~\cite{PHOBOS:2010eyu,PHENIX:2000owy} and will be referred to as "the combinatoric method". The second is primarily guided by the LHC Run 2 \xexe and Run 3 \pbpb analyses by the \cms Collaboration~\cite{CMS:2019gzk, CMS:2024ykx} and will be referred to as "the closest-match method". Aspects of the analysis which are common to both methods will be discussed jointly, while the approach-specific methods will be introduced and explained separately.

\subsection{Common analysis components}
A set of \intt calibrations is implemented to ensure data quality and to optimize hit reconstruction.
First, hot, cold, and dead channels are identified from the data, accounting for approximately 4\% of all \intt channels, and are excluded from the analysis in both real data and simulation. Second, only hits associated with the triggered beam bunch crossing are retained. Third, an analog-to-digital conversion (ADC) calibration is applied, mapping the recorded \intt hit amplitude to the predefined threshold setting of the corresponding comparator.

After extracting \intt hits, hits are clustered using an adjacency graph. Two \intt hits are considered adjacent if and only if they share the same coordinate in $z$ and have touching edges in the azimuthal ($\phi$) direction. Ideally, these clusters represent the full extent of the energy deposition from a charged particle passing through an \intt layer and contain information about its location, timing, size, and energy. A cluster ADC threshold is applied to exclude single-hit clusters with the minimal hit ADC value, which are predominantly noise. Clusters originating from collision-produced particles typically contain at least one strip with a higher hit ADC value. Additionally, unphysically large clusters, which likely result from a rare saturation issue where an \intt chip only read out a portion of its recorded hits, are removed. The impact of this saturation issue is minimal, affecting approximately $10^{-5}$ of the total reconstructed tracklets.

The definition of tracklets is identical between the two analysis approaches. Clusters originating from a particle track associated with the event vertex exhibit small differences in pseudorapidity, azimuthal angle, and angular separation. These variables are defined as $\DEta = \eta_{\text{inner}} - \eta_{\text{outer}}$, $\DPhi = \phi_{\text{inner}} - \phi_{\text{outer}}$, and $\DR=\sqrt{(\Delta\eta)^2 + (\Delta\phi)^2}$.
Here, $\eta_{\text{inner(outer)}}$ and $\phi_{\text{inner(outer)}}$ represent the pseudorapidity and azimuthal angle of the cluster in the inner (outer) layer of the \intt, calculated with respect to the event vertex unless specified otherwise. Both vertex reconstruction and tracklet counting utilize the fact that tracklets associated with particles originating from the event vertex produce a coincidence peak in the $\DEta$, $\DPhi$, and $\DR$ distributions.

The two analyses diverge in the event vertex determination, tracklet reconstruction and counting, and the correction factors, which are discussed separately in the following subsections. The position of the event vertex in the x and y directions is determined by the beamspot, defined as the average vertex position in the transverse plane over multiple events, and the vertex position along the z-axis, referred to as the z-vertex hereafter, which is determined on an event-by-event basis. The transverse vertex position varies by $\mathcal{O}(100)\mum$, which is several orders of magnitude smaller than the length scale of the radial distance between the \intt ladder and the beamline. In contrast, the z-vertex variation is significantly larger, approximately 9.4\cm. 
Determining the z-vertex on an event-by-event basis is crucial for accurately establishing the tracklet kinematics. This is achieved with a reconstruction resolution of 0.17\cm for the selected centrality interval as determined in simulation.

\subsection{The combinatoric method}
Within the combinatoric method, the beamspot position is determined using two techniques: the iterative quadrant search and the two-dimensional tracklet filling method:  

\begin{itemize}
    \item \textbf{The iterative quadrant search}. A bounded square region in the transverse plane, centered at the origin of the \sphenix coordinate system, is defined. The four corners of the square serve as initial beamspot candidates. For each tracklet candidate, $\phi_{\text{inner}}$, $\phi_{\text{outer}}$, $\DPhi$, and the distance-of-closest-approach (DCA) are calculated relative to the candidate position. 
    The correlations between DCA and $\phi_{\text{inner}}$, and between $\DPhi$ and $\phi_{\text{inner}}$ are fitted to a constant function, utilizing the fact that deviations from a horizontal line indicate an incorrect beamspot assumption. 
    The quadrant containing the candidate with the smallest fit uncertainty is selected for the next iteration, where its four corners serve as new beamspot candidates. This iterative process continues until the search window is reduced to a size comparable to the spatial resolution of \intt strips.
    \item \textbf{The 2-dimension tracklet filling method}. The trajectories of cluster pairs are populated into a finely segmented two-dimensional histogram. The beamspot position is obtained as the mean value of the bins with a content exceeding 70\% of the maximum bin content.
\end{itemize}
The iterative quadrant search determines the beamspot position along the x- and y-axes as ($x^{\text{iterative}}_{\text{beamspot}}, y^{\text{iterative}}_{\text{beamspot}}) = (-0.019, +0.230)$\cm, while the 2D tracklet filling method yields ($x^{\text{2-d}}_{\text{beamspot}}, y^{\text{2-d}}_{\text{beamspot}}) = (-0.025, +0.215)$\cm. The final beamspot position used in the analysis of the combinatoric method is determined as the arithmetic mean of both measurements.
  
The $\phi_{\text{inner}}$ and $\phi_{\text{outer}}$ values are updated based on the measured beamspot. For each cluster pair in a given event, a trapezoidal probability distribution is constructed by projecting all possible trajectories that could connect the clusters while crossing the strip acceptances of both along the beam axis. The trapezoidal distributions from cluster pairs are combined to form a stacked distribution. The z-vertex is determined as the average of the mean values from Gaussian functions fitted to the distribution with different fit ranges.

\begin{figure}[t!]
  \centering
  \includegraphics[width=0.5\textwidth]{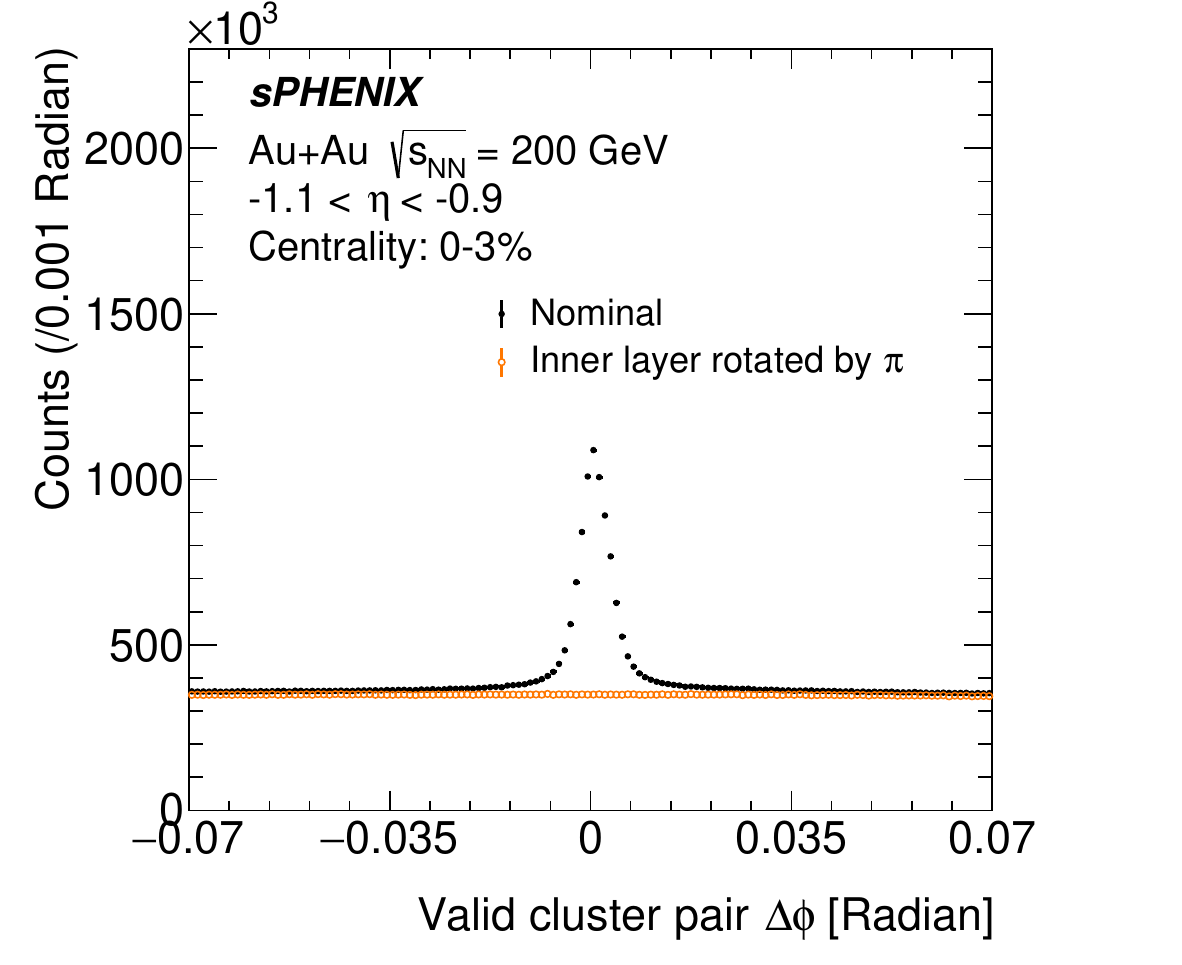}
  \caption{The $\Delta\phi$ distributions for valid cluster pairs in a representative $\eta$ and centrality interval. The solid black circles indicate the distribution of all valid cluster pairs, while the open orange circles represent the estimated combinatorial background, obtained by rotating the inner-layer ladders by $\pi$ around the beam axis.}
  \label{fig:PHOBOS_deltaphi}
\end{figure}

The $\eta$ values of clusters are recalculated with respect to the estimated z-vertex. Candidate cluster pairs are formed from all possible associations of clusters in the inner layer with those in the outer layer. Clusters are not restricted to a single cluster pair. Valid cluster pairs are defined as those whose trapezoidal projections intersect the reconstructed vertex position. The combinatorial background is estimated by repeating the same pairing procedure after rotating the inner layer by $\pi$ in $\phi$, as illustrated in Figure~\ref{fig:PHOBOS_deltaphi}. This background component is then subtracted from the number of valid cluster pairs. The contribution of combinatorial background is strongly dependent on cluster multiplicity, ranging from 88.4\% in the most central events to 27.0\% in the most peripheral events. The number of reconstructed tracklets within a given $\eta$ and centrality interval is obtained by counting the remaining valid cluster pairs with $|\DPhi|<\text{0.026}$ radians. This $\DPhi$ cut effectively suppresses contributions from particles with $p_T < 50\mev/\text{c}$, including those from two-photon production of dilepton pairs.   

The corrections, derived from simulation, account for the difference between the number of charged hadrons produced in collisions and the number of combinatorial-corrected tracklets, considering acceptance and efficiency losses. For each centrality interval, an $\eta$-dependent correction factor is defined as the ratio of the average number of reconstructed tracklets to the average number of charged hadrons. Within the measured centrality intervals, the correction factors range from 0.8 to 1.0 for $|\eta|<1.1$. These corrections are then applied to the data to obtain the measured \dndeta.

\subsection{The closest-match method}
\label{sec:analysis-closest}
The beamspot position is determined using the DCA-$\phi$ fitting method, closely following the approach developed by the \cms experiment~\cite{Miao:2007zz}. This approach takes advantage of the fact that, for tracks originating from a beamspot at $(x_0,y_0)$, the DCA to the origin exhibits a sinusoidal pattern with respect to the $\phi$ coordinate of the point-of-closest-approach (PCA) to the origin ($\phi_{\text{PCA}}$):
\[
\text{DCA}(\phi_{\text{PCA}})=R_0\cos(\phi_{\text{PCA}}-\phi_0)
\]
where $R_0 = \sqrt{x_0^2 + y_0^2}$ is the radial coordinate of the beamspot, and $\phi_0 = \arctan\big(\frac{y_0}{x_0}\big)$ represents its $\phi$ coordinate. By fitting the sinusoidal pattern formed by the tracklet DCA and $\phi_{\text{PCA}}$, the parameters $R_0$ and $\phi_0$ are extracted.
The determined beamspot position in the x- and y-axis ($x^{\text{DCA-$\phi$ fit}}_{\text{beamspot}}, y^{\text{DCA-$\phi$ fit}}_{\text{beamspot}})=(-0.029, +0.225)\cm$.

The event z-vertex is reconstructed using the following procedure. First, clusters in the inner layer are paired with clusters in the outer layer, with cluster $\phi$ values calculated relative to the beamspot coordinates. For each pair, a z-vertex candidate is constructed, with its boundaries defined by linearly extrapolating the edges of the cluster pair to a line pointing along the longitudinal direction through the beamspot. The candidate range is then divided into fine segments and filled into a one-dimensional histogram. This histogram is fitted with a Gaussian plus a constant offset, and the mean of the Gaussian component is taken as the reconstructed z-vertex position. 

\begin{figure}[t!]
  \centering
  \includegraphics[width=0.5\textwidth]{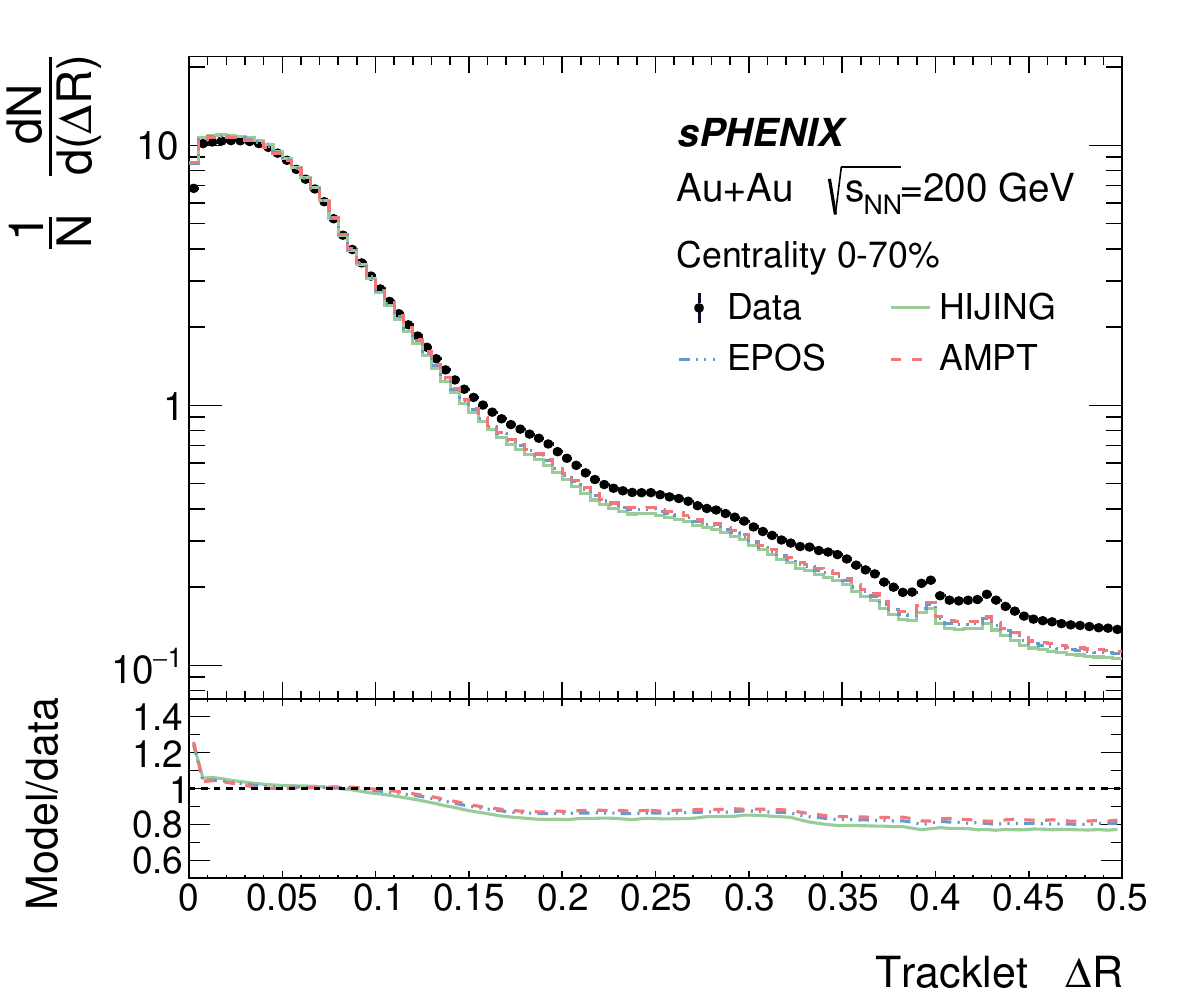}
  \caption{The $\DR$ distributions for reconstructed tracklets, normalised by the total number of tracklets. The spectrum in data (black circle) is compared to simulations generated with \hijing, \epos, and \ampt.}
  \label{fig:recotkldR_approach1}
\end{figure}

Cluster $\eta$ and $\phi$ values are updated relative to the measured beamspot and z-vertex coordinates. Within an event, tracklets are formed by pairing one cluster in the inner barrel with one in the outer barrel. Combinations with $\DR$ less than 0.5 are retained and sorted by $\DR$. This selection criterion ensures that tracklets originating from primary particles are preserved with minimal inefficiency while limiting contamination from secondary particles and combinatorial background to at most 10\%. If multiple matches exist for a cluster, the pair with the smallest $\DR$ is selected to form the final set of reconstructed tracklets. Figure~\ref{fig:recotkldR_approach1} presents the $\DR$ distributions for reconstructed tracklets in data, compared to fully simulated events generated by \hijing, \epos, and \ampt.

Similar to the combinatoric method, the reconstructed tracklet spectrum is corrected to the hadron-level event definition in simulation. In the closest-match method, three multiplicative correction factors are derived and applied. The first compensates for discrepancies between the actual \intt geometry and the geometry implemented in simulations and reconstruction. In simulations, hits are both generated and reconstructed using the same assumed geometry, whereas in real data, minor deviations persist due to detector misalignments that have yet to be implemented in the software geometry. This correction factor is determined by randomly populating tracklets in both data and simulation to estimate the acceptance in the $\eta$ and z-vertex phase space based on tracklet counts and is obtained as the ratio of the tracklet counts from the randomly populated distributions. The correction ranges between 0.9 and 1 for $|\eta|<1.1$.
The second correction, denoted as the $\alpha$ factor, accounts for tracklet reconstruction inefficiencies and is defined as the ratio of primary charged hadrons in simulation to uncorrected reconstructed tracklets. This factor is determined as a function of $\eta$ and z-vertex and applied separately for each centrality interval to correct for efficiency and fake rate variations. The third correction addresses acceptance variations due to shifts in the z-vertex position, which impact tracklet reconstruction near acceptance edges. It is obtained by comparing the total number of tracklets per z-vertex bin to those reconstructed in valid $\alpha$-factor regions and is applied as a function of $\eta$.
The $\alpha$ factor and the acceptance correction are applied jointly, with values ranging between 0.8 and 1.2 for $|\eta|<1.1$.

%% file: systematic.tex
\section{Systematic Uncertainties}
Systematic uncertainties affecting the \dndeta measurement are evaluated. These include statistical limitations in simulations, uncertainties in cluster and tracklet selection criteria, variations across run segments, model dependence of the correction factors, contributions from secondary particles originating from weak decays, effects of detector misalignment, and uncertainties in centrality determination and Glauber model parameters. These systematic uncertainties apply to both methods unless explicitly stated. Table~\ref{tab:uncsummary} provides a summary of the magnitude of systematic uncertainties affecting \dndeta for both analysis approaches, where the reported ranges detail the smallest and largest uncertainties measured in each $\eta$ bin, over all centrality classes. The correlated and uncorrelated uncertainties of the weighted average result, detailed in Section~\ref{sec:results}, are also presented.

\begin{itemize}
    \item The uncertainty in the correction factors arises from the limited statistics of the simulation. To quantify this uncertainty, the correction factor is varied within its statistical uncertainty. The modified correction is then applied to the uncorrected number of tracklets, and the resulting variation in the \dndeta is evaluated and quoted as the systematic uncertainty. 
    \item The cluster ADC criterion influences the signal-to-background ratio of clusters. Its impact is evaluated by adjusting the cut level.
    \item The effect of \intt chip saturation is studied and evaluated by performing the analysis without the cluster $\phi$-size cut. 
    \item The sensitivity of the correction factors to tracklet selection criteria is examined by varying the nominal selection criteria in both approaches, accounting for potential effects on the signal-to-background ratio of reconstructed tracklets. 
    \item The impact of potential variations from different machine and data-taking conditions and the stability of detector components is assessed by segmenting the data sample and repeating the analysis for each segment. The baseline \dndeta distribution is obtained from the first segment, while the largest deviation observed across all segments is quoted as a systematic uncertainty.
    \item The dependence of correction factors on the choice of event generator is studied by comparing results from \hijing, \epos, and \ampt, which model particle production and kinematics differently. These differences affect the ratio of generated hadrons to reconstructed tracklets, leading to variations in correction factors.
    \item Decays of strange particles can generate multiple clusters, potentially resulting in double or multiple counting in the \dndeta measurement. The sensitivity of the correction factors to secondary particles from weak decays is evaluated using a \hijing sample with a 40\% enhancement in strangeness. This enhancement fraction is chosen as a conservative yet realistic estimate, reflecting the differences in strange particle composition between the \hijing and \epos event generators.
    \item In the combinatoric method, the impact of detector misalignment is assessed by introducing random displacements of \intt ladders within $\pm250\,\mum$ in simulation. In the closest-match method, this effect is addressed through a correction factor, detailed in Section~\ref{sec:analysis-closest}, that accounts for geometric discrepancies between data and simulation. Due to this correction factor, this systematic uncertainty is not separately evaluated in the closest-match method.
    \item Uncertainties in \npart values are estimated by applying standard variations in the MC Glauber model and centrality determination, including the variation on the nucleon-nucleon cross-section, geometric parameters in the Glauber model, and the effect of the uncertainty in the event selection effciency to shift the events into other centrality intervals. For the centrality intervals used in this analysis, the uncertainties range from 0.6\% for 0–3\% central events to 14.9\% for 65–70\% peripheral events.
\end{itemize}

The uncertainty associated with the cluster ADC selection is the dominant source for both approaches. The approximately 3\% difference between the two methods in this uncertainty arises from their differing sensitivities to combinatorial backgrounds. The combinatoric method, which reconstructs all possible cluster pairs, is more susceptible to these backgrounds, as changes in the number of available clusters have a greater impact on the number of reconstructed tracklets. In contrast, the closest-match method, which selects only the cluster pair with the smallest $\DR$, is less affected by such variations. 

\begin{table}[H]
    \centering
    \caption{Systematic uncertainties from various sources are reported, with the range indicating the minimum and maximum uncertainty magnitudes across all $\eta$ bins and centrality intervals. Additionally, the correlated and uncorrelated uncertainties of the weighted average result, as detailed in Section~\ref{sec:results}, are provided.}
    \begin{adjustbox}{width=1\textwidth}
    \begin{tabular}{l c c}
        \toprule[1pt]
        \textbf{Source} & \textbf{The combinatoric method [\%]} & \textbf{The closest-match method [\%]} \\
        \midrule[1pt]
        Simulation statistics & 0.1--0.6 & 0.2--0.9 \\
        Cluster ADC selection & 3.8--8.8 & 2.8--5.4 \\
        Cluster $\phi$-size selection & $<\text{0.1}$ & $<\text{0.2}$ \\
        Tracklet reconstruction criteria & 0.7--1.2 & $<\text{1.7}$ \\        
        Machine and detector stability & $<\text{1.0}$ & 0.1--1.6 \\
        Model dependence & 0.5--5.7 & 1.6--3.8 \\
        Secondaries & $<\text{2.6}$ & $<\text{3.2}$ \\
        Detector misalignment & 0.5--0.9 & -- \\\cmidrule(rl){2-3}
        \textbf{Total} & 4.1--10.3 & 3.5--6.9 \\
        \midrule[1pt]
        Correlated uncertainty in the weighted average result & \multicolumn{2}{c}{3.5\%--7.9\%} \\
        Uncorrelated uncertainty in the weighted average result & \multicolumn{2}{c}{$<\text{0.9}\%$} \\
        Total uncertainty in the weighted average result &\multicolumn{2}{c}{3.5\%--7.9\%} \\
        \bottomrule[1pt]
    \end{tabular}
    \label{tab:uncsummary}
    \end{adjustbox}
\end{table}

%% file: result.tex
\section{Results}
\label{sec:results}
Results from the combinatoric method and the closest-match method are statistically combined. Systematic uncertainties in the two methods are classified based on their correlation coefficients: those with a correlation coefficient greater than 0.1, such as uncertainties from simulation statistics, cluster ADC and $\phi$-size selections, machine and detector stability, model dependence, and secondary contributions, are treated as fully correlated. In contrast, uncertainties with a correlation coefficient below 0.1, such as those arising from tracklet reconstruction criteria, are considered uncorrelated. 
The weighted average of the two approaches, $\bar{X}$, and the uncorrelated uncertainty on the weighted average result, $\sigma_X$, are computed as:  
\begin{equation}
\begin{split}
\bar{X} \pm \sigma_X & = \frac{ w_{\text{Comb.}} X_{\text{Comb.}} + w_{\text{C.-m.}} X_{\text{C.-m.}}}{w_{\text{Comb.}} + w_{\text{C.-m.}}} \pm \left(w_{\text{Comb.}} + w_{\text{C.-m.}} \right)^{-1/2}, \\ 
 w_{\text{Comb.}} & = \frac{1}{(\sigma_{X,\text{Comb.}})^2}\quad \text{,}\quad w_{\text{C.-m.}} = \frac{1}{(\sigma_{X,\text{C.-m.}})^2}, \\
\end{split}
\label{eq:weiavg-uncorrunc}
\end{equation}

where \(X_\text{Comb./C.-m.}\), \(\sigma_{X,\text{Comb./C.-m.}}\), and $w_{\text{Comb./C.-m.}}$ represent the value, uncorrelated uncertainty, and the weight reported by the two methods. The weights vary with centrality and pseudorapidity, with an average weight ratio of $w_{\text{C.-m.}} / (w_{\text{Comb.}} + w_{\text{C.-m.}})$ = 0.83.
The correlated uncertainty on the weighted average result, $\bar{s}$, is calculated:  
\begin{equation}
\bar{s} = \sqrt{\sum_k \bigg[\frac{w_{\text{Comb.}}}{w_{\text{Comb.}}+w_{\text{C.-m.}}}(s_{\text{Comb.}})_k + \frac{w_{\text{C.-m.}}}{w_{\text{Comb.}}+w_{\text{C.-m.}}}(s_{\text{C.-m.}})_k\bigg]^2},
\label{eq:corrunc}
\end{equation}
where $(s_{\text{Comb.}})_{k}$ and $(s_{\text{C.-m.}})_{k}$ denote the uncertainties from the combinatoric method and the closest-match method (indexed by $k$), respectively. The total uncertainty of the weighted average result is obtained by:  
\begin{math}
\sigma_{\text{total}} = \sqrt{\bar{s}^2 + \sigma_X^2}.
\end{math} 

The combination strategy used in this measurement closely follows procedures outlined by the Particle Data Group and the \cms experiment~\cite{ParticleDataGroup:2024cfk, CMS:2011aqh}, specifically to account for dominant correlated uncertainties in the two analysis methods, where Equations~\ref{eq:weiavg-uncorrunc} and \ref{eq:corrunc} are applicable. Notably, the combined uncertainty has a modest increase compared to the method with the smaller uncertainty in some instances. In contrast, the use of a best limit unbiased estimator~\cite{5f88760e-8167-31ed-9022-4d8c4dafd76a} is seen to decrease the overall uncertainty but cause the central value of some bins to fall outside the mean values defined by the individual methods. The approach adopted in this paper may not be universally applicable, and careful consideration is required when extending it to other measurements with different correlation structures.

\begin{figure}[H]
  \centering
  \includegraphics[width=0.85\textwidth]{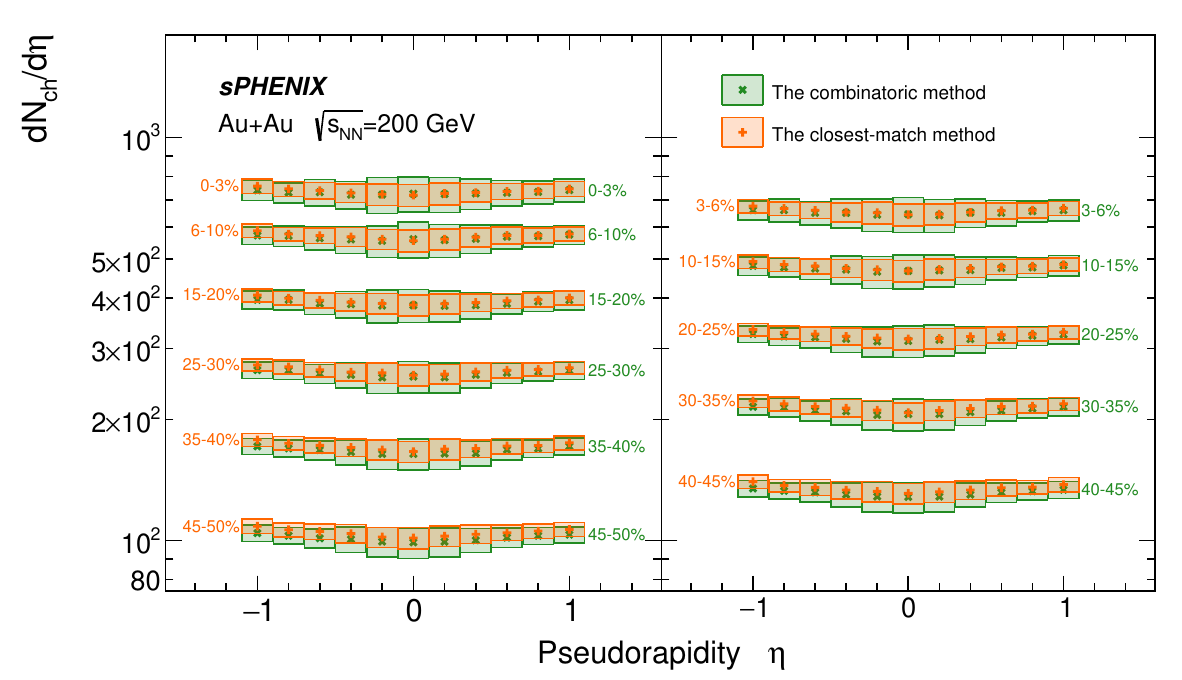}\\
  \includegraphics[width=0.85\textwidth]{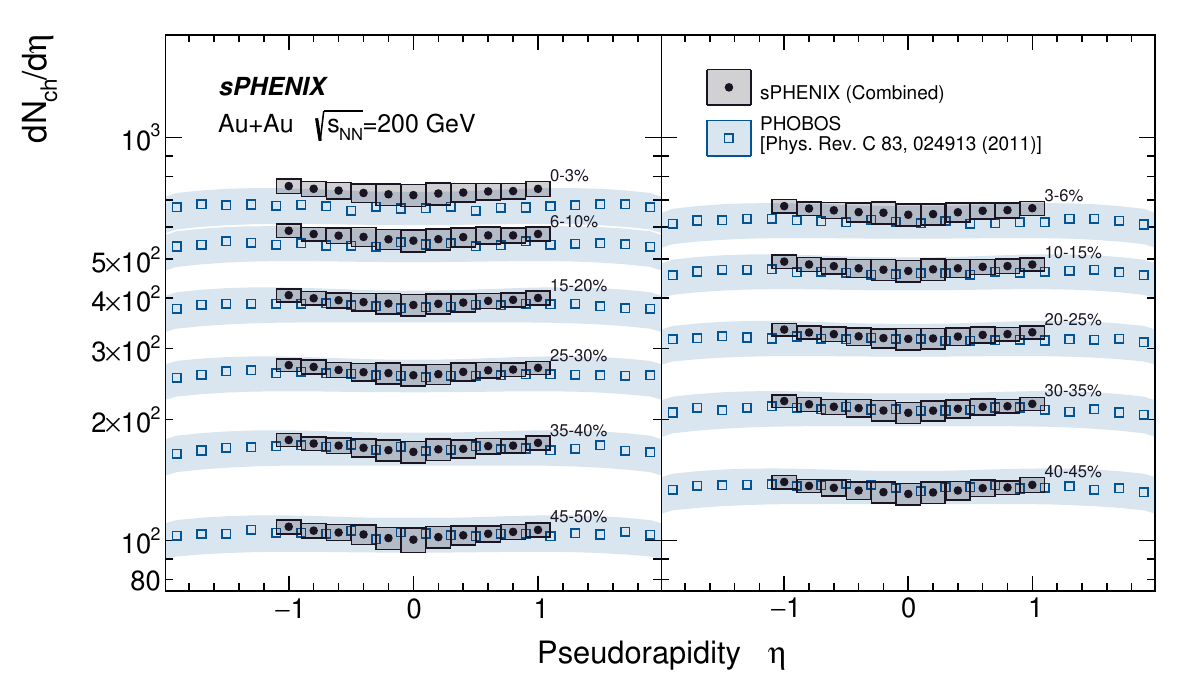}
  \caption{Top: corrected \dndeta measurements from the combinatoric method (green solid crosses and green transparent filled boxes) and the closest-match method (orange solid pluses and transparent filled boxes) as a function of $\eta$ over the measurement range $|\eta|<1.1$, spanning from 0–3\% central to 45–50\% mid-central \auau events. The vertical extent of each box represents the total systematic uncertainty for each analysis method, obtained by summing all sources in quadrature under the assumption that individual uncertainty sources are independent and uncorrelated. Bottom: \dndeta measurements from the combination of both analysis methods (black solid circles and transparent filled boxes) as a function of $\eta$ over the range $|\eta|<1.1$, covering centralities from 0–3\% to 45–50\% in \auau collisions. The vertical size of each box represents the full combined uncertainty, calculated using the procedure outlined above. For comparison, \phobos measurements in the same centrality classes are shown as blue open squares~\cite{PHOBOS:2010eyu}, with the filled area representing the total uncertainty.} 
  \label{fig:dndeta-allcent}
\end{figure}

Figure~\ref{fig:dndeta-allcent} presents \dndeta as a function of $\eta$ for various centrality intervals, showing results from both approaches as well as the combined result compared with measurements from \phobos~\cite{PHOBOS:2010eyu}. The results from both approaches are consistent within their respective uncertainties, which are obtained by summing all sources in quadrature under the assumption that individual uncertainties are independent and uncorrelated. The combined result remains in agreement with the \phobos measurements across the $\eta$ range covered in this analysis, with uncertainties in the \sphenix measurement calculated using the procedure described above.

\begin{figure}[t!]
  \centering
  \includegraphics[width=0.5\textwidth]{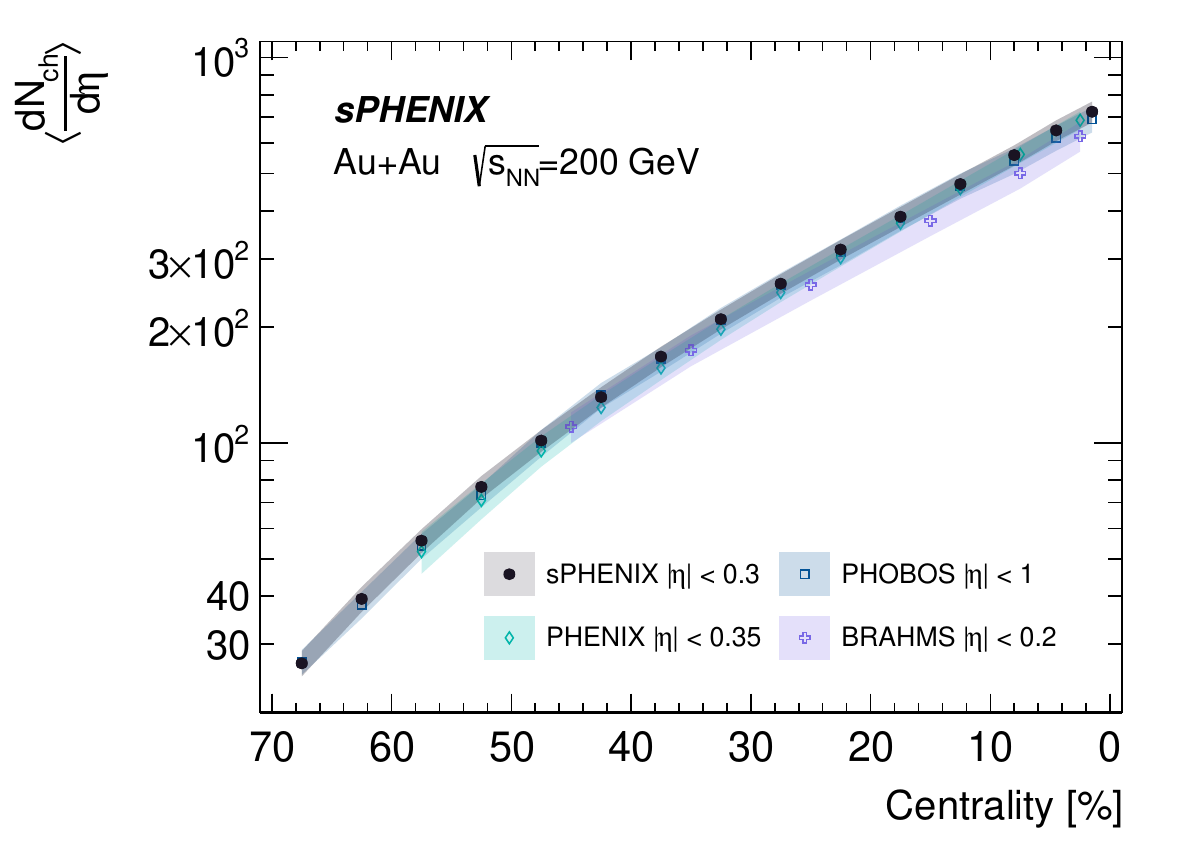}~
  \includegraphics[width=0.5\textwidth]{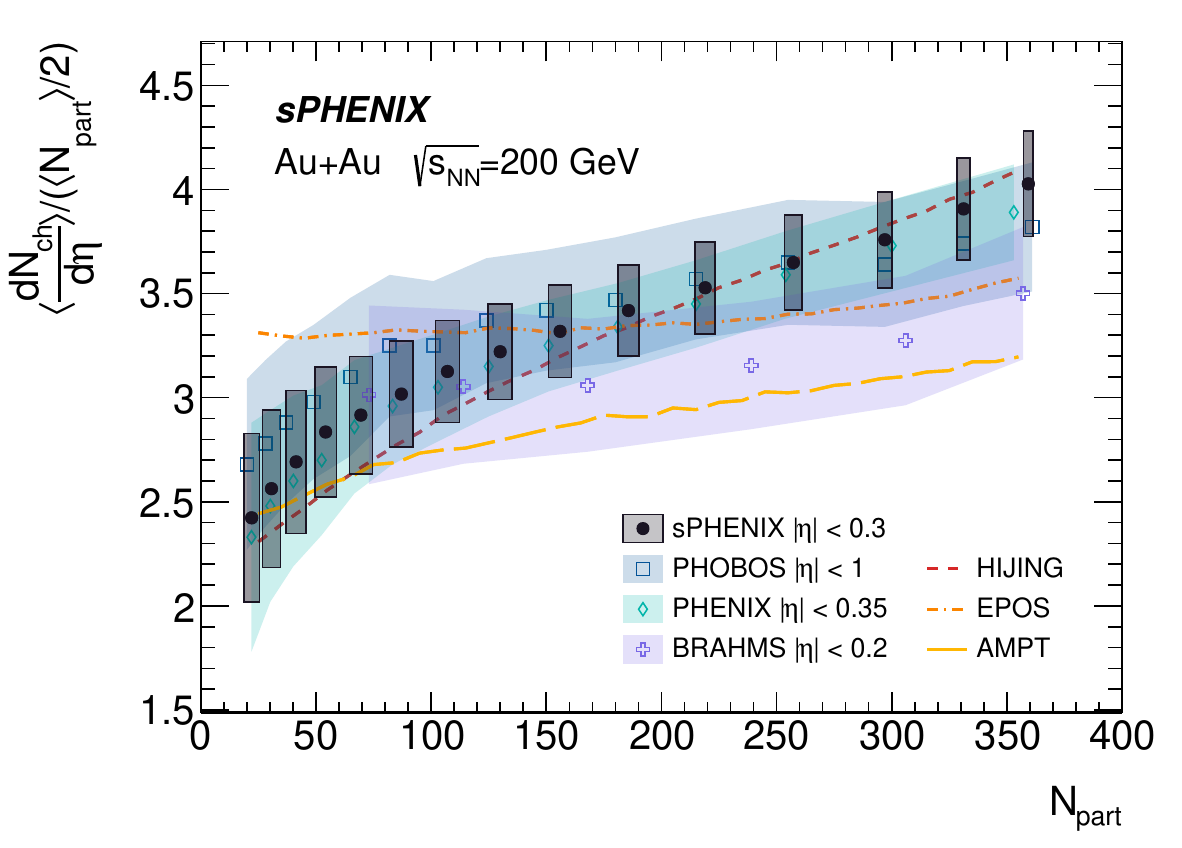}
  \caption{The \sphenix results are shown as black solid circles with transparent filled band/boxes indicating total uncertainties, while data from \phobos (blue open squares)~\cite{PHOBOS:2010eyu}, \phenix (cyan open diamonds)~\cite{PHENIX:2004vdg}, and \brahms (purple open crosses)~\cite{BRAHMS:2001llo} are presented for comparison. The colored bands represent the total systematic uncertainty for each presented measurement. Left: \sphenix combined \dndeta at mid-rapidity, averaged over \(|\eta| < 0.3\), as a function of event centrality. Right: the average \dndeta at mid-rapidity, normalised by the number of participant pairs, as a function of the number of participating nucleons.}
  \label{fig:dndeta-eta0avgnpart}
\end{figure}

The left panel of Figure~\ref{fig:dndeta-eta0avgnpart} presents \dndeta at mid-rapidity, averaged over $|\eta|<0.3$, as a function of centrality class, demonstrating that charged hadron multiplicity increases with collision centrality. The charged hadron density for the most central 3\% of events (0–3\% centrality bin) is measured to be $\dndeta|_{|\eta| < 0.3} = 723.4 \pm 45.3$.
Results from this analysis (black solid circles), obtained by combining two analysis methods as described above, are compared with previous RHIC measurements, including \phobos (blue open squares), \phenix (cyan open diamonds), and \brahms (purple open crosses). The shaded color bands represent the full systematic uncertainty. The \sphenix results are consistent with those from \phobos, \phenix, and \brahms within uncertainties across all centrality intervals. 

The right panel of Figure~\ref{fig:dndeta-eta0avgnpart} presents the average \dndeta per participant pair, $N_{\text{part}}/2$, as a function of $\langle N_{\text{part}}\rangle$, to evaluate the relationship between particle production and the number of participating nucleons. The \sphenix data exhibit the characteristic non-linear increase, with a steeper rise in more peripheral collisions compared to central collisions. The measured $\dndeta|_{|\eta|<0.3}/(N_{\text{part}}/2$) values range from approximately $2.4$ to $4.0$ charged hadrons per participant pair on average, over the \npart range covered in this analysis.
The \sphenix results are consistent with those from \phobos, \phenix, and \brahms within uncertainties across all centrality bins considered. This observable is further compared to predictions from the \hijing, \epos, and \ampt MC heavy-ion event generators, with results indicating that the \sphenix measurement is most compatible with \hijing in both magnitude and \npart dependence. 

The midrapidity charged hadron multiplicities, the number of participants, and the charged hadron multiplicities normalised to the number of participant pairs, $\langle$\npart$\rangle/2$, are summarized in Table~\ref{tab:results}. 

\begin{table}[t!]
    \centering
    \caption{Summary of the midrapidity charged hadron multiplicities, the number of participants, and the charged hadron multiplicities normalised to the number of participant pairs $\langle$\npart$\rangle/2$.}
    \onehalfspacing
    \begin{tabular}{c c c c}
        \toprule[1pt]
        \textbf{Bin} & \textbf{$\frac{dN_{\text{ch}}}{d\eta}\big|_{|\eta|<0.3}$} & $\langle$\textbf{\npart}$\rangle$ & $\frac{\dndeta|_{|\eta|<0.3}}{\langle N_{\text{part}}\rangle/2}$ \\
        \midrule[1pt]
        0\%-3\% & $723.4\pm45.3$ & $359.3\pm2.1$ & $4.0\pm0.3$ \\
        3\%-6\% & $646.9\pm40.2$ & $331.2\pm2.9$ & $3.9\pm0.2$ \\
        6\%-10\% & $558.1\pm33.7$ & $297.0\pm3.2$ & $3.8\pm0.2$ \\
        10\%-15\% & $469.5\pm28.5$ & $257.3\pm3.8$ & $3.6\pm0.2$ \\
        15\%-20\% & $386.3\pm23.0$ & $219.0\pm4.3$ & $3.5\pm0.2$ \\
        20\%-25\% & $317.4\pm18.6$ & $185.7\pm4.6$ & $3.4\pm0.2$ \\
        25\%-30\% & $258.9\pm15.2$ & $156.0\pm5.0$ & $3.3\pm0.2$ \\
        30\%-35\% & $209.3\pm12.3$ & $130.0\pm5.2$ & $3.2\pm0.2$ \\
        35\%-40\% & $167.4\pm10.2$ & $107.1\pm5.2$ & $3.1\pm0.2$ \\
        40\%-45\% & $131.4\pm8.0$ & $87.1\pm5.1$ & $3.0\pm0.3$ \\
        45\%-50\% & $101.3\pm6.6$ & $69.5\pm5.0$ & $2.9\pm0.3$ \\
        50\%-55\% & $76.8\pm5.2$ & $54.2\pm4.7$ & $2.8\pm0.3$ \\
        55\%-60\% & $55.7\pm3.9$ & $41.4\pm4.4$ & $2.7\pm0.3$ \\
        60\%-65\% & $39.3\pm2.9$ & $30.7\pm3.9$ & $2.6\pm0.4$ \\
        65\%-70\% & $26.8\pm2.0$ & $22.1\pm3.3$ & $2.4\pm0.4$ \\
        \bottomrule[1pt]
    \end{tabular}
    \label{tab:results}
\end{table}

%% file: conclusion.tex
\section{Conclusion}
This paper presents the measurement of charged hadron multiplicity per unit pseudorapidity, \dndeta, using field-off data from Run 2024 in \auau collisions at $\sqrtsnn=200\gev$, collected with the sPHENIX detector. Results are reported as a function of pseudorapidity, $\eta$, within $|\eta|<1.1$ across multiple centrality intervals. In the $3\%$ most central collisions, \dndeta for $|\eta|<0.3$ is measured to be $723.4\pm45.3$. The \dndeta strongly increases with collision centrality, and the normalised \dndeta per participant pair exhibits a mild rise with the number of participating nucleons, \npart. 
The \sphenix measurement, which combines two analysis methods, provides full $2\pi$ azimuthal coverage at mid-rapidity and presents a characterization of charged particle multiplicity as a function of $\eta$. These results are in agreement with previous measurements from \phobos, \phenix, and \brahms, further supporting the broader sPHENIX physics program.
